# A PSO Strategy of Finding Relevant Web Documents using a New Similarity Measure

Ramya C[1*], Shreedhara K S[2]

[1*]Research Scholar, Department of Studies in Computer Science and Engineering, UBDTCE, Davanagere, Karnataka, INDIA
[2]Professor, Department of Studies in Computer Science and Engineering, UBDTCE, Davanagere, Karnataka, INDIA

**Abstract**

In the world of the Internet and World Wide Web, which offers a tremendous amount of information, an increasing emphasis is being given to searching services and functionality. Currently, a majority of web portals offer their searching utilities, be it better or worse. These can search for the content within the sites, mainly text – the textual content of documents. In this paper a novel similarity measure called SMDR (Similarity Measure for Documents Retrieval) is proposed to help retrieve more similar documents from the repository thus contributing considerably to the effectiveness of Web Information Retrieval (WIR) process. Bio-inspired PSO methodology is used with the intent to reduce the response time of the system and optimizes WIR process, hence contributes to the efficiency of the system. This paper also demonstrates a comparative study of the proposed system with the existing method in terms of accuracy, sensitivity, F-measure and specificity. Finally, extensive experiments are conducted on CACM collections. Better precision-recall rates are achieved than the existing system. Experimental results demonstrate the effectiveness and efficiency of the proposed system.

## 1. Introduction

Information retrieval (IR) is a crucial technology in accessing the vast amount of data present on today's World Wide Web. It is also one of the critical techniques in knowledge management. Developments in IR began already in the pre-computer era and continued today. The evaluation of IR has been an integral part of IR development. It is the moving force which influences the advances in the field. Several annual conferences that deal exclusively with IR testifies to the dynamism of the area. Effectiveness evaluation depends on the notion of relevant and non-relevant documents concerning a query. The key to evaluating an IR system is to devise an evaluation collection composed of a document collection, a group of questions and the relevance judgements for the pairs of documents and queries.

Searching for any other differently structured data will not be very successful or is not currently supported [2]. Mathematical documents contain a precisely tremendous amount of data. Yet, often when WIR systems are assessed for their performance, numerous relevant documents found to be missed [3]. Furthermore, user is the ultimate judge of the document relevance because user would judge whether a document is completely relevant or not to his query. A related problem is that the document collection is ever-changing and the user queries are static. To address the above issues, we propose a novel similarity measure which addresses the effectiveness issue by measuring the similarity between documents thereby helps to retrieve most relevant documents fending off non-relevant ones. We develop a system that accepts the user query as input and uses PSO methodology for optimisation of WIR process, hence contributes to the efficiency of the system reducing response time, and achieves good precision and recall rates in WIR.

The main objective of the current study is to explore a WIR which could retrieve most relevant documents to the users with a reduced response time hence devoting to effectiveness and efficiency of the system. More precisely, we aim at come up with a novel similarity measure which gauges the similarity between the documents and helps to rank the results. In addition, we aim at optimising WIR process using PSO in order to retrieve the documents with a plausible amount of time. We focus on examining the effectiveness of the novel similarity measure and efficiency of the system developed, using various evaluation metrics concerned to IR, with the intent of examining on a huge standard corpus CACM using various queries.

[*]*Corresponding Author:  e-mail: cramyac@gmail.com,*
*Tel-+91-9164508779*
ISSN 2320-7590

Habiba Drias et. al. [1] developed two novel algorithms PSO1-IR and PSO2-IR to improve the performance of the WIR and showed the significant reduction in response time relatively to the traditional approaches. Cosine similarity function is used to match between query and the documents. However, there is a scope to improvise the similarity measure which is used in the system so that it helps to retrieve still more relevant documents from the repository without missing any. This is considered as existing system and has been implemented and used for comparative study along with the proposed system. The next section describes related work on how PSO is used in WIR and similarity functions used in WIR by others. Novel similarity metric and proposed systems are highlighted in section 3. Experimental results analysis is described in section 4. We conclude our work in Section 5.

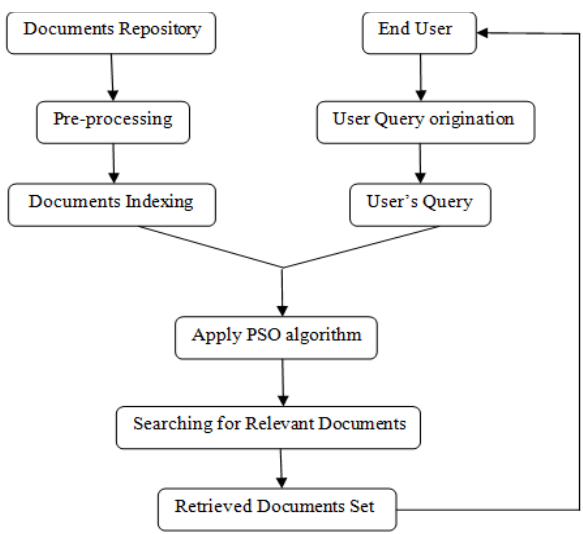

**Figure 1.** Proposed architecture

## 2. Related Work

Swarm Intelligence (SI) is a contemporary optimisation tool used to solve numerous optimisation problems from various fields. Over the past few years, SI methods like PSO are used to optimize the WIR process and seemed to be very successful. Habiba Drias et al. [4] demonstrated the parallelism of PSO to achieve high quality relevant documents with a reduced response time. The two novel parallel PSOs ultimately show quick responses to the queries executed than the traditional algorithms [5, 6, 7]. However, in case of huge documents collection parallel PSO attains low similarity values compared to exact algorithm [8] but provides considerably better response time than it.

Pamba et al. [9] studied and observed the evaluation of the efficiency of hardness and softness of an innovative approach to often occurring pattern growth based fuzzy PSO for collecting web documents. The formal methods of Fuzzy C-Means (FCM) and K-Means fail in the context of random initialization and local minima hook-ups. To overcome these limitations, bio-inspired mechanisms such as genetic algorithm; ACO and PSO was utilized to enhance the K-means and FCM clustering. The main contributions of the new method were three fold. Firstly, in its ways to spontaneously found active cluster numbers, cluster centroids, and swarms for the bio-inspired fuzzy PSO. Secondly, to overcome the problems of the existing approaches the yield of fuzzy overlapping clusters through the FCM objective function is made. Thirdly, the methodology argued in this study reduced the separate elements from the search space and thus engaged all associations with search query as semantic conditionally relatable sets.

Irfan et al. [10] presented a key framework of the IR process using evolutionary computation and sure of the fundamental models which were being used. Different unnatural methods such as ACO, PSO, and genetic algorithm were utilized for enhancing the data by considering different sets of algorithms which were somewhat better than traditional computing methods. It comprised a summary of computation and how optimization assistances to recover improved outcomes using different techniques.

Khennak et al. [11] proposed the application of the APSO technique to successfully resolve the issue of Query Expansion (QE) in WIR. Unlike previous studies, the authors introduced innovative modelling of QE that targeted to found the appropriate expanded query from a set of extended query candidates. However, owing to the significant number of potential extended query candidates, it was tremendously difficult to create the best one through conventional hard computing techniques. Thus, they suggested the QE problem as a combinatorial optimization issue and addressing it through APSO. Suggested APSO aimed at QE with MEDLINE, the world Web's largest medical library are thoroughly assessed.

Djenouri et al. [12] proposed improvements in the field of data mining to solve the primary document IR problem. Useful knowledge was first predicted by considering data mining methods, and then swarms utilize this knowledge to predict the entire space of documents logically in the suggested approach. The authors examined two data mining methods in the pre-processing step. The initial one intended to split the documents collection into similar clusters by utilizing the K-means algorithm; then the second one extracted the most closed frequent terms on each cluster already formed using the DCI Closed algorithm. For solving the phase, Bees Swarm Optimization was used to calculate the group of documents honestly. The suggested method has been assessed on recognized collections such as ACM, Text Retrieval Conference, Web docs, etc. and it linked to modern data mining, bio-inspired in addition to further documents IR based methodologies. The result showed that the recommended method improves the quality of returned documents significantly, with the processing time compared to advanced approaches.

Sharma et al. [13] presented a combined technique for IR from MRI images of the brain, and the technique depends on the Artificial Intelligence and k-means clustering. The process of extraction of feature includes the Grey level co-occurrence. Extracted feature for Fuzzy Inference System includes morphological operator, thresholding, and Watershed segmentation for detection of a brain tumor. The recommended technique was employed to identify affected brain part and tumor size from MRI image with the support of MATLAB R2013b.

Hao et al. [14] presented a retrieval structure which was based on lattice. This construction demonstrated a document in addition to a user query in a concept space revealed on fuzzy formal concept analysis utilized a semantic index to arrange documents and ranks documents by considering learned concept. Experiments were performed on the text collections achieved from the SMART IR system.

Kuper et al. [15] studied the influence of recovery approach on the specificity of memory content recovered in the course of understanding and reminiscence processes. In two trials, participants implemented a recognition memory inclusion task in which they had to differentiate identically reiterated and re-colored forms of study items from new items.

Khatiwada et al. [16] suggested a novel example of IR techniques to maintain bug localization tasks in software schemes. These methods, associated with Normalized Google Distance and Point-wise Mutual Information, exploit the concurrent patterns of code terms in the system to show the hidden textual semantic dimensions that additional techniques often were unsuccessful in capturing. The objective was found to be optimistic semantic similarity connections amongst source code as well as bug reports.

Hand et al. [17] presented that the weighted sum of recall and optimism gives the F-measure, weights given based on the association technique being utilized. This reformulation exposed that the F-measure has a significant conceptual weakness: the relative spotlight allocated to recall and precision that includes in the issue and the researcher, but not of the precise linkage technique being utilized. They suggested alternative measures which do not grieve from this fundamental defect.

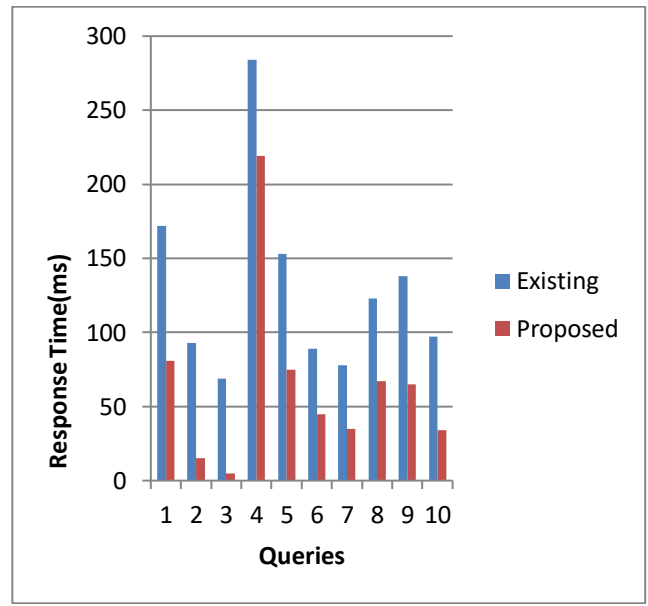

**Figure 3.** Response time

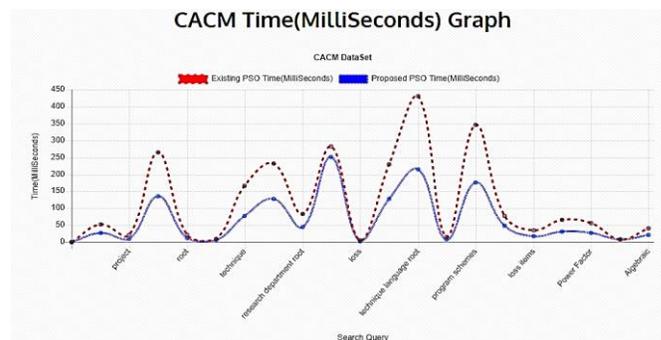

**Figure 4.** Snapshot of a plot of response time for different queries.

Similarity function plays a great role in retrieving relevant documents to the user query. It just reflects how similar a document is for the issued query by assigning the rank to it. Significant work has been carried out on similarity measure. Anna Huang [18], in his work, compared and analysed Euclidean distance, Cosine similarity, Jaccard coefficient, Pearson correlation coefficient and averaged Kullback-Leibler divergence for their effectiveness in text clustering. R.Subhashini et al. [19] found that Cosine and Jaccord yield comparable effectiveness while Euclidean is ineffective to help to fetch the more relevant documents. Yung-Shen Lin et al. [20] proposed a new measure to gauge the similarity between the two documents $d_1 = <d_{11}, d_{12}, \ldots, d_{1m}>$ and $d_2 = <d_{21}, d_{22}, \ldots, d_{2m}>$ called SMTP (Similarity Measure for Text Processing).

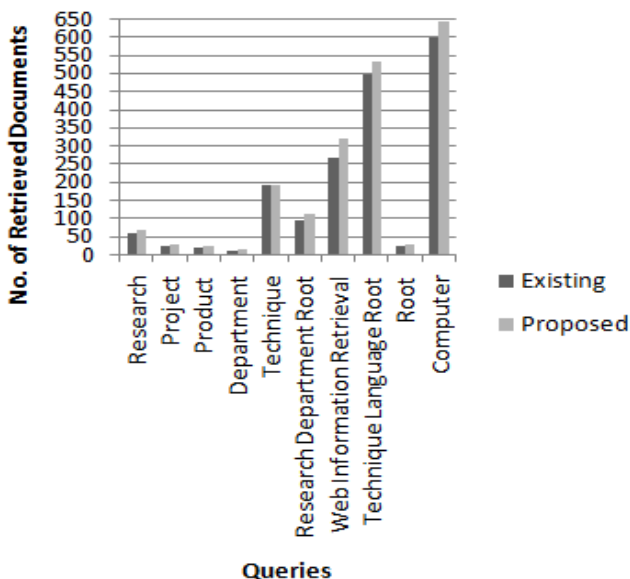

**Figure 2.** Proposed v/s existing system on number of documents retrieved for the different queries.

### 3. Web Information Retrieval

WIR is the application of IR to the web. In traditional IR, consumers specify queries, in particular, query language, demonstrating their information requirements. The scheme chooses the set of documents in its group that appear the most significant to the query and offers them to the user. Users can refine their questions for enhancing the answer. In the environment of a web, user aims are not stationary and steady as they typically are in traditional IR. IR systems were initially established to handle the enormous amount of information. Several universities, corporate and libraries now utilize IR systems to provide access to books, journals and other documents. Decades of research in WIR were fruitful in emerging as well as refining methods that are exclusively word-based. With the arrival of the web new sources of information turn out to be accessible, one of them being the hyperlinks amongst documents and records of user behaviour. The objective of WIR system in Figure 1 is to satiate user's information requirement. Unfortunately, description of user information need is not simple. User's every so often do not know obviously about the information need. A query is only a vague and partial report of the information need.

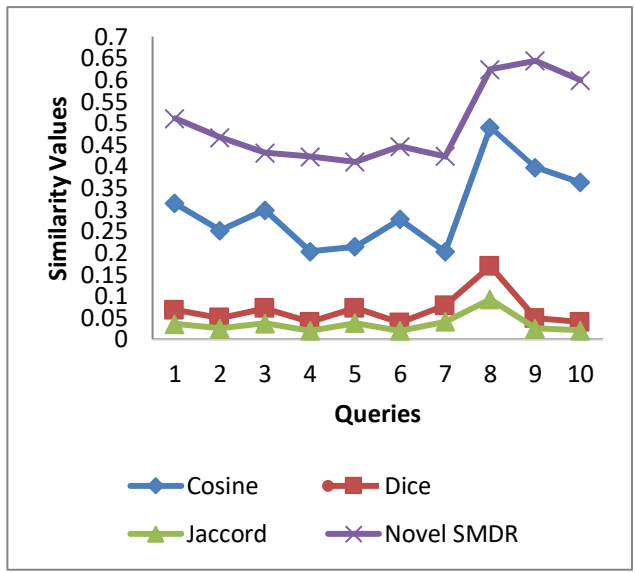

**Figure 5.** Performance of Similarity Measures

WIR is ultimately a task of discovering appropriate data from a more significant collection of unstructured data. WIR process runs in the background, uses a huge documents repository and the user query indicating user needs as input and retrieves the most relevant documents at the top as output. A document can be a structured data, text, video, image, sound, musical scores, DNA sequences, etc. A document is normally depicted by a set of keywords/terms contained in it. The user queries and documents must be represented as per a model. Vector space model is the extensively used one where in vectors of term weights depict both documents and queries. Vector space encloses all the terms that the system come across and is constructed during indexing process. Term weight designates the significance of the term in the query or in the document. For the issued query, Similarity values for the documents are computed using a similarity measure. Several techniques are used to rank between documents based on the computed similarity. The top ranked documents are deemed relevant to the query and presented as output. The proposed architecture can be seen in Figure 1.

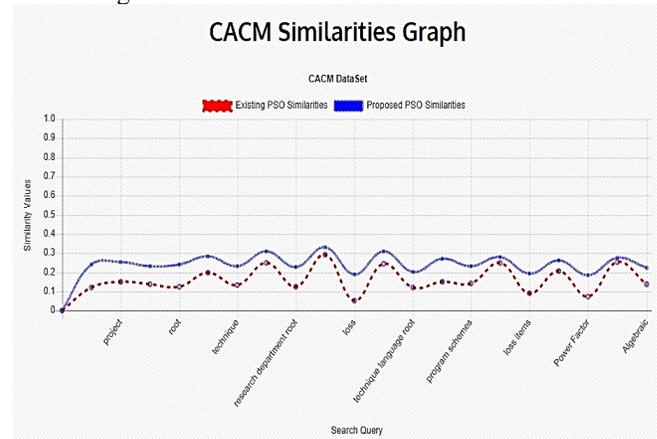

**Figure 6.** Snapshot of a plot of similarity values of SMDR on different queries.

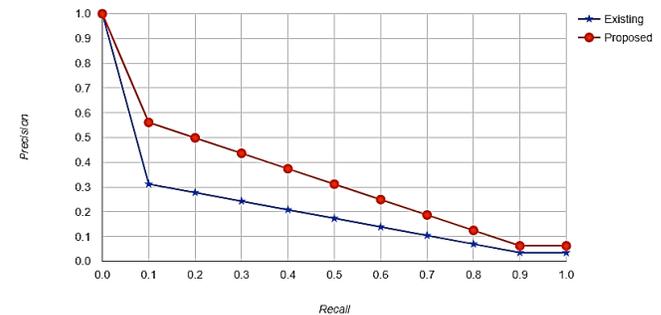

**Figure 7.** Recall and Precision graph for the query "research"

*3.1 Preprocessing:* Documents preprocessing has a noticeable impact on effectiveness and efficiency of the documents retrieval. It is worth to preprocess the text contained in documents to obtain the specific terms to store in the index. Its procedure is as follows. Firstly, the words in the text of documents are identified and considered as index terms. Space is the words separator. Punctuation marks are usually removed. Numbers are vague without a context, hence usually disregarded as good index terms. Hyphenated words are broken. No matter with the case of letters. All the text is converted into either upper or lower case. This is called lexical analysis. Subsequently, stop words are eliminated by removing the most appearing words like "is", "of", "i", "over", "if" and so forth. Stop word removal considerably reduces the size of the index. Next, stemming of the words by removing suffixes and prefixes is carried out. For instance, the word *"reflect"* is the stem for the variants *reflected, reflecting, reflection* and *reflections.* Stemming contributes in decreasing the size of the index by forming stems of the words. The mentioned text operations can easily be implemented but the care has to be taken on each operation as they have sever effect on document retrieval.

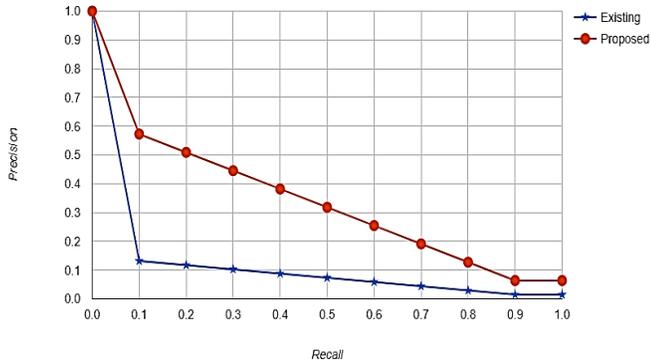

**Figure 8.** Recall and Precision graph for the query "product"

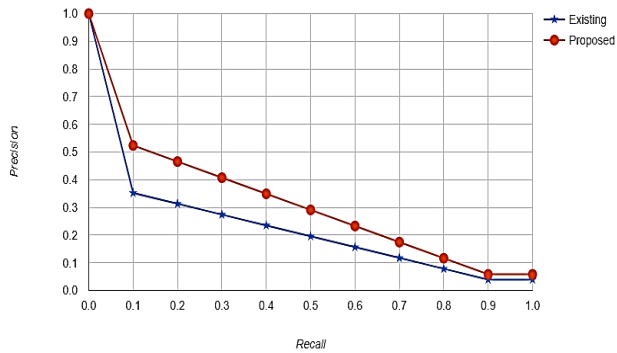

**Figure 9.** Recall and Precision graph for the query "research root product"

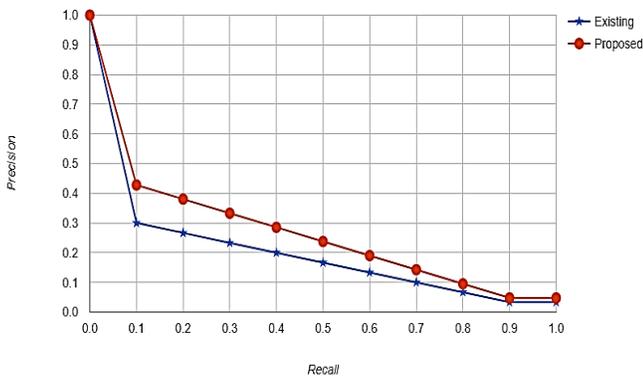

**Figure 10.** Recall and Precision graph for the query "information retrieval"

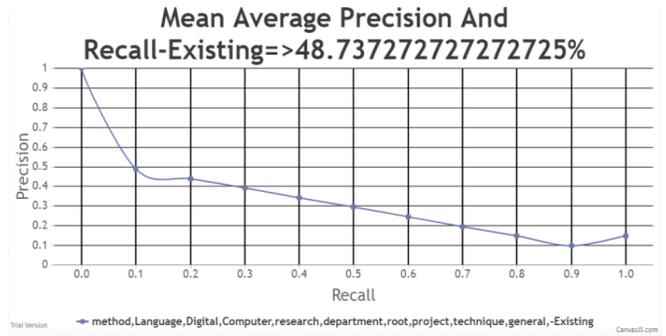

**Figure 11.** Mean Average Precision-Recall curve of existing system.

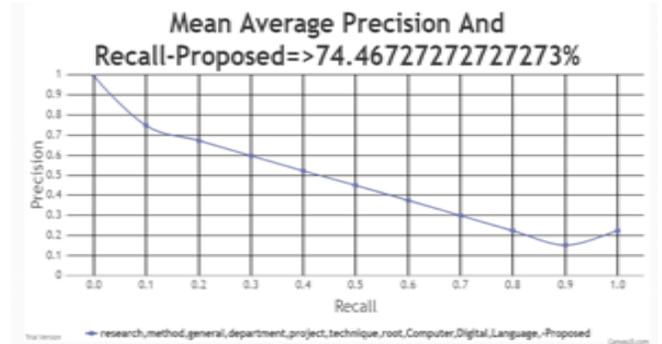

**Figure 12.** Mean Average Precision-Recall curve of proposed system.

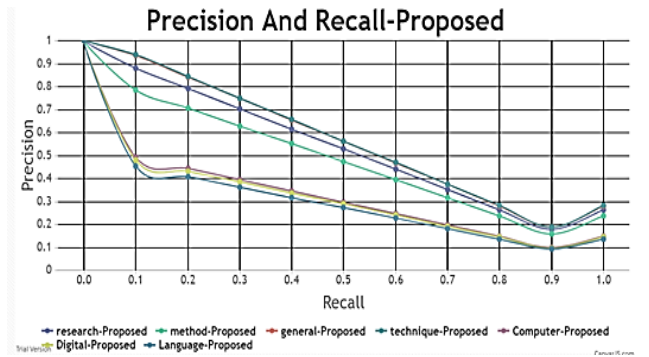

**Figure 13.** Precision-Recall curve of proposed system for multiple queries.

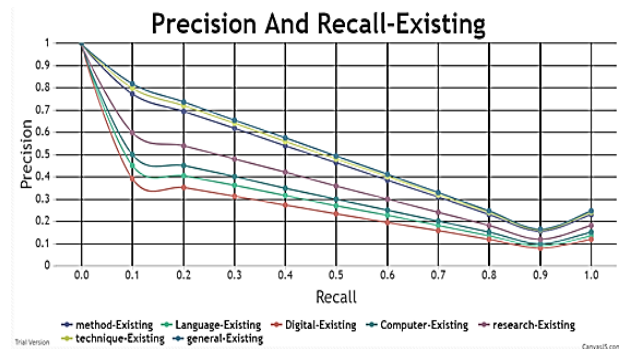

**Figure 14.** Precision-Recall curve of existing system for multiple queries.

*3.2 Novel Similarity Function:* Similarity measure is a significant parameter used to rank the similar documents and thereby helps to retrieve the most relevant documents from the documents collection. Based on the desirable properties to qualify as a similarity metric, a novel similarity measure, called SMDR (Similarity measure for Documents Retrieval), for a document d = <$t_1,t_2,t_3, ... ,t_n$> (containing maximum of 'n' terms) and a query q = <$qt_1, qt_2, qt_3, ... , qt_h$> (involving 'h' number of terms at most) defines a function NSim(q,t) as follows:

$$\text{NSim(q,t)} = \frac{\sum_{i=0}^{n-1} tc_i qhc_i q_i t_i}{\sqrt{\sum_{1=0}^{n-1}(\frac{tc_i}{qhc_i})^2} + \sqrt{\sum_{1=0}^{n-1}(q_i)^2} * \sqrt{\sum_{1=0}^{n-1}(t_i)^2}}$$

Where, NSim= Novel Similarity, q=query, t=terms, tc=terms count and qhc=query hit terms count. This novel SMDR has been used in proposed system to match between the query and documents. It serves to be the better measure for providing similarity values with respect to the vector of query terms thus helps ranking the documents.

*3.3 Particle Swarm Optimization:* PSO is a swarm intelligence approach inspired by the behaviour models of birds flocking, originally developed by J. Kennedy and R.C. Eberhart 1995. It's a scenario of group of birds (swarm) searching for a food in an area randomly which are unaware of the place where food is located. But the birds know how distant the food is from the present location. The entire collection of documents is a search space. Each document is considered as a particle and so the position of the particle is its document identifier. The search process is scanning the documents one by one is assumed as a particle moving from one position to another position in the search space. WIRS accepts the user query and the indexed documents as input. Once the documents are pre-processed, PSO algorithm is applied and the novel SMDR is used as fitness function to find out the relevance between the documents.

*Steps:*
1. Initialize the indexing for the documents.
2. Assign docs to indexwriter

$$Indexwriter_k = docs_k$$

3. Compute maximum no. of terms in each and every index.

$$Maxterms_k = indexwriter_k.counttrems()$$

4. Repeat the steps 2 and 3, $\forall\ k = 1,2,....n$.
5. Initialize the Clustering.
6. Compute the Cluster,

$$Cluster_h = \begin{cases} Maxterms_h, & Maxterms_h < Maxterms_{h+1} \\ Maxterms_{h+1}, & \&\ Otherwise \end{cases}$$

7. Repeat the step 6 $\forall\ i = 1,2,.....n$
8. Initialize the particles by assigning Clusters, for each clusters i to m.
9. Compute the velocity

$$\begin{aligned}&\text{randpart=randval ,0}\\&<\text{randval}<((\text{insf}+1)*2)/2\\&\text{randglob = randval ,0}\\&< randval < ((\text{insf} + 1) * 2)/2\end{aligned}$$

$$vel_{bc} = insf * vel_{bc} + randpart * (part_{bc} - newvel_{bc}) + randglob * (bss - newvel_{bc})$$

$$newvel_{bc} = (newvel_{bc} + vel_{bc}) \bmod (totdocscount) + 1$$

10. Update the best solution for particle best cluster.

$$newvel_{bc} > f(part_{bc}) \rightarrow part_{bc} = newvel_{bc}$$
$$f(part_{bc}) > f(bss) \rightarrow bss = part_{bc}$$

11. Repeat the steps 9 and 10 for every particle.

Where,
- *k* and *h* as loop counters.
- *n* and *m* are total number of documents.
- *indexwriter* to store all the documents to construct the index
- *maxterm* stands for total number of terms in index.
- *bc* is the best cluster for maxterms.
- *randpart* and *randglob* indicate the confidence coefficients. *randpart* is for particle's personal best value and *randglob* is neighbourhood best value.
- *insf* makes the particle to move with the same velocity and direction is called as inertia factor.
- *vel* is velocity for the particle *part.*
- *bss* stands for the best global solution for the whole swarm.
- *part[bc]* is the best solution for particle.
- *f* is nothing but novel SMDR used to calculate fitness values in PSO.

However, having PSO implemented, it is observed that PSO fails to keep its potential to seek out a global optimum solution and convergence rate turns relatively slower as well. Hence further part of the research work focuses on addressing these drawbacks of PSO. PSO will be hybridised with Simulated Annealing technique which has the capability to converge to the global optimum solution causing the retrieval of most relevant documents.

*3.4 Datasets Used:* CACM is a collection of 3204 HTML documents involving articles abstracts of ACM journal published in around 20 years. These HTML documents are containing totally 6768 terms in them while the average document size is relatively equal to 2 Kb. Table 2 shows the characteristics of datasets used.

## 4. Experimental Results and Discussion

The proposed novel similarity measure SMDR with PSO optimization is implemented in Java and have gone through numerous experiments.

*4.1 Experimental Setup:* The empirical parameters of PSO such as number of particles and number of iterations are set initially to provide better quality results. These are set to 20 and 50 respectively for CACM data. Systems are Tested for large number of various queries.

**Table 1.** Various evaluation measures

| Query | Accuracy | | Sensitivity | | Specificity | | F-Measure | |
|---|---|---|---|---|---|---|---|---|
| | Existing | Proposed | Existing | Proposed | Existing | Proposed | Existing | Proposed |
| 1 | 99.0637 | 99.7815 | 97.7273 | 97.7273 | 99.0823 | 99.8101 | 74.1300 | 92.4731 |
| 2 | 98.9700 | 99.8752 | 60.0000 | 60.0000 | 99.2760 | 100.1887 | 47.6190 | 88.2353 |
| 3 | 89.2634 | 88.9825 | 88.8889 | 89.1813 | 89.3082 | 88.9588 | 63.8655 | 63.3437 |
| 4 | 99.2509 | 100.3121 | 55.5560 | 66.6667 | 99.3740 | 100.4069 | 29.4118 | 35.6523 |
| 5 | 97.8464 | 98.5019 | 100.000 | 100.000 | 97.8165 | 98.4810 | 56.0510 | 64.7059 |
| 6 | 96.3483 | 97.2846 | 72.9630 | 72.9630 | 98.5003 | 99.5228 | 77.1037 | 81.9127 |
| 7 | 93.7890 | 93.8202 | 85.0299 | 86.2275 | 94.2707 | 94.2377 | 58.7992 | 59.2593 |
| 8 | 93.7578 | 94.0075 | 74.3405 | 74.3405 | 96.6631 | 96.9501 | 75.6098 | 76.3547 |
| 9 | 98.8764 | 99.8439 | 53.1250 | 53.1250 | 99.3380 | 100.3153 | 48.5714 | 87.1795 |
| 10 | 94.7566 | 95.6929 | 56.0554 | 56.0554 | 98.5935 | 99.6226 | 65.8537 | 70.1299 |

*4.2 Proposed versus Existing systems:* Proposed system retrieved considerably more relevant documents than the existing system [1]. It clearly shows the effect of SMDR on documents retrieval. Figure 2 shows the plot of the number of documents retrieved by both existing and proposed system. Hence SMDR contributes to the effectiveness of WIR. A soon as the user query is issued, the time taken by the system to search and retrieve the relevant documents is its response time. Here, the objective is to reduce it as much as possible. Figure 3 shows significant reduction in the response time of the proposed system than the existing one. Figure 4 is the snapshot of the same. Hence contributing to the efficiency of WIR process.

**Table 2.** Characteristics of datasets

| Datasets | CACM |
|---|---|
| No. Of Terms | 6768 |
| No. Of Documents | 3,204 |
| Average Document Size (Bytes) | 2K |

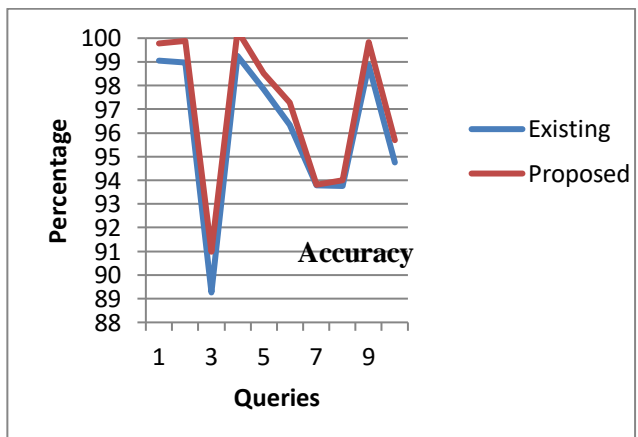

**Figure 15.** Accuracy

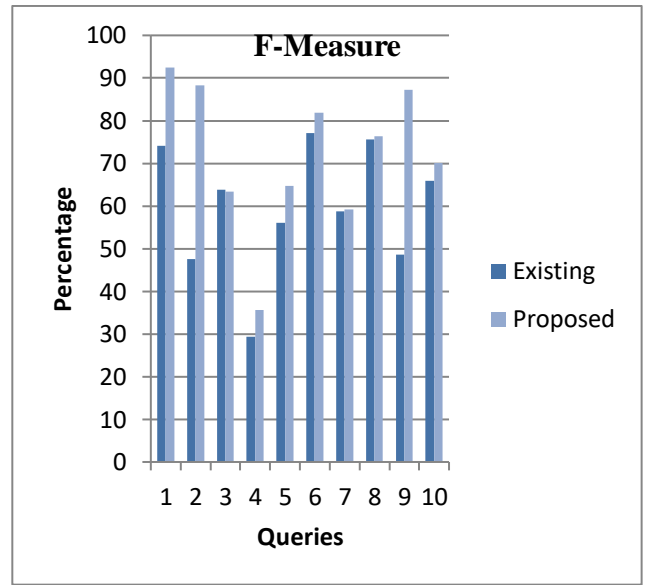

**Figure 16.** F-Measure

*4.3 Evaluation of Novel Similarity measure SMDR :* In order to evaluate novel SMDR, Average relevancy of each set of document for a single query was calculated using Cosine, Dice and Jaccord similarity coefficients as fitness functions which are popularly used in literature. Figure 5 shows graphical representation of performance of Cosine, Dice, Jaccord and Novel SMDR functions. The curve of novel SMDR clearly shows that the proposed system offers good performance exploiting the power of novel SMDR. Figure 6 shows the snapshot of a graph of similarity values of SMDR for a set of queries in percentage.

*4.4 Precision-Recall Curves:* Here, the recall and precision are used as retrieval evaluation measures. Recall is the quantity of the relevant documents that are actually retrieved and Precision is the amount of the retrieved documents that are really relevant. Figures 7-10 show the precision-recall curves for the mentioned queries (single). The P/R curves clearly show the better rates for proposed than the existing system. Figures 13 and 14 demonstrate the output of Precision-Recall curves for multiple queries for both proposed and existing systems respectively.

In a document repository, the arithmetic mean of average precision values at several recall levels for an individual user query to measure retrieval accuracy is known as Mean Average precision (MAP). Figures 11 and 12 display the snapshot of plot for MAP at various recall points drawn for both existing and proposed system respectively. Undoubtedly, proposed system has obtained better MAP values than existing one.

*4.5 Other Evaluation measures:* Accuracy is the portion of true results among the entire documents in the repository. It is calculated using the expression as in (1). Figure 15 demonstrates that the proposed system is more accurate than the existing system. Similarly sensitivity and specificity are calculated by equations (2) and (3) respectively. F- Measure is another performance measure used to evaluate the efficiency of the system. It takes into consideration both precision and recall together and provides single measurement for a system. The mathematical expression of the same is given in (4). Figure 16 presents the F-measure values in percentage for various queries. The superiority of proposed system can be seen over existing one [1]. Table 1 exhibits various evaluation metrics used.

Accuracy= ((TP+TN)/ (TP+TN+FP+TN));      (1)
Sensitivity= (TP/ (TP+FN))      (2)
Specificity= (TN/ (TN+FP))      (3)
F-Measure=2*((Precision*Recall)/ (Precision+Recall))   (4)

Where,
TP- True Positive
TN- True Negative
FP- False Positive
FN- False Negative

## 5. Conclusions and Future Work

This paper presents a novel similarity function SMDR for retrieving better-quality in the sense highly relevant documents from the repository. Its effectiveness is compared with the popular and widely used similarity functions Cosine, Dice and Jaccord. Results show that novel SMDR contributes significantly to the effective documents retrieval than the existing ones. The proposed system uses PSO algorithm for optimising WIR process which helps in quickly retrieving the relevant documents for the user with considerably reduced response time. PSO uses SMDR as its fitness function to arrive with the superior performance than existing system [1]. Proposed system gives significant precision-recall rates. The system is evaluated for its accuracy, sensitivity, specificity and F-measure values.

As for as future work is concerned, the system will be tested with RCV1 datasets which is a huge repository of exactly 8,04,414 documents. Considering the gigantic size of WWW, experiments on CACM collection having 3204 html pages is not sufficient. So a large scale of datasets RCV1 will be used to prove that the proposed method is generic. Furthermore, PSO algorithm will be hybridised to overcome its limitation of premature convergence.

**Acknowledgement**

Authors gratefully acknowledge the Department of studies in Computer Science and Engineering, UBDTCE, Davanagere for providing all the necessary facilities to accomplish the research work and also express sincere appreciation to anonymous reviewers for the many helpful comments and criticism.

**Biographical notes**

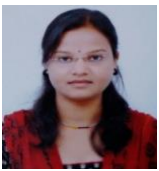

**Ramya C** has received M.Tech. in Computer Science and Engineering from Davanagere University in 2011 and pursuing Ph.D. in the Department of studies in Computer Science & Engineering, University B.D.T College of Engineering, Davanagere from Visvesvaraya Technological University, Belagavi, Karnataka, India. Her research interests lie in Information Retrieval, Data Mining, Soft Computing and Neural Networks.

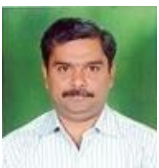

**Shreedhara K S** has received M.Tech in System Analysis from NITK, Surathkal, Mangalore University in 1997 and Ph.D. from Manipal University, Karnataka, India in 2008. He is a professor in the Department of studies in Computer Science & Engineering, University B.D.T College of Engineering, Davanagere, Karnataka, India. His research interest includes Data Mining, Machine Learning, Pattern Recognition, Soft Computing, Artificial Intelligence, Computer Graphics and Image and Video Processing.